\documentclass[conference]{IEEEtran}
\IEEEoverridecommandlockouts

\usepackage{cite}
\usepackage{amsfonts}
\usepackage{amsthm}
\usepackage{algorithmic}
\usepackage{graphicx}
\usepackage{textcomp}
\usepackage[table]{xcolor}
\usepackage{algorithm}
\usepackage{multirow}
\usepackage{amsmath}
\usepackage{subcaption}
\usepackage{placeins} %
\usepackage{bm}
\usepackage{makecell}
\usepackage{pgfplots}
\usepackage{balance}
\usetikzlibrary{positioning} 
\usepackage{caption} 
\usepackage{adjustbox}
\usepackage{xspace}
\usepgfplotslibrary{groupplots}
\pgfplotsset{compat=newest}
\usepackage[inline]{enumitem}

\newcommand{\ourschemelo}{LoByITFL\xspace}

\newcommand{\Totalclient}{\ensuremath{n}}
\newcommand{\Byzantine}{\ensuremath{e}}
\newcommand{\Colluding}{\ensuremath{t}}
\newcommand{\Dropout}{\ensuremath{s}}
\newcommand{\partition}{\ensuremath{m}}

\newcommand{\Dataset}{\ensuremath{D}}

\newcommand{\model}{\ensuremath{\bm{w}}}

\newcommand{\dimension}{\ensuremath{d}}
\newcommand{\globalIteration}{\ensuremath{g}}
\newcommand{\localIteration}{\ensuremath{c}}
\newcommand{\learningRate}{\ensuremath{\eta}}

\newcommand{\trustscore}{\ensuremath{\mathrm{TS}}}
\newcommand{\modelupdate}{\ensuremath{\mathbf{u}}}
\newcommand{\realnormalmodelupdate}{\ensuremath{\tilde{\mathbf{u}}}}
\newcommand{\normalmodelupdate}{\ensuremath{\bar{\mathbf{u}}}}

\newcommand{\sumone}{\ensuremath{\Sigma_1}}
\newcommand{\sumtwo}{\ensuremath{\mathbf{\Sigma}_2}}

\newcommand{\random}{\ensuremath{\lambda}}

\newcommand{\approdegree}{\ensuremath{k}}

\newcommand{\randomvector}{\ensuremath{\mathbf{r}}\xspace}

\DeclareMathOperator{\ReLU}{ReLU}
\DeclareMathOperator{\AGG}{AGG}

\newtheorem{thm}{Theorem}

\newtheorem{proposition}{Proposition}
\newtheorem{remark}{Remark}

\def\BibTeX{{\rm B\kern-.05em{\sc i\kern-.025em b}\kern-.08em
    T\kern-.1667em\lower.7ex\hbox{E}\kern-.125emX}}
    
\begin{document}
\pagestyle{plain}
\title{LoByITFL: Low Communication Secure and Private Federated Learning\\
\thanks{This project has received funding from the German Research Foundation (DFG) under Grant Agreement Nos. BI 2492/1-1 and WA 3907/7-1. This is the author's version of the work. The definitive Version of Record will be published at the International Workshop on Secure and Efficient Federated Learning In Conjunction with ACM AsiaCCS 2025 (FL-AsiaCCS’25), https://doi.org/10.1145/3709023.3737688.}}

\author{\IEEEauthorblockN{Yue Xia, Maximilian Egger, Christoph Hofmeister, Rawad Bitar}
\IEEEauthorblockA{
School of Computation, Information and Technology, Technical University of Munich, Munich, Germany \\
\{yue1.xia, maximilian.egger, christoph.hofmeister, rawad.bitar\}@tum.de}
}
\maketitle

\begin{abstract}
Privacy of the clients' data and security against Byzantine clients are key challenges in Federated Learning (FL). Existing solutions to joint privacy and security incur sacrifices on the privacy guarantee. We introduce \ourschemelo, the first \emph{communication-efficient} information-theoretically private and secure FL scheme that makes no sacrifices on the privacy guarantees while ensuring security against Byzantine adversaries. The key components are a small and representative dataset available to the federator, a careful modification of the FLTrust algorithm, and the one-time use of a trusted third party during an initialization period. We provide theoretical guarantees on the privacy and Byzantine resilience, as well as experimental results showing the convergence of \ourschemelo.
\end{abstract}

\maketitle
\thispagestyle{plain}

\section{Introduction}
Federated learning (FL)~\cite{mcmahan2017communication} emerged as a promising paradigm that enables training machine learning models on private data owned by multiple clients. Following an iterative structure, per iteration, the federator sends the current global model to the clients, who compute local model updates based on their individual training data and return the results to the federator, updating the global model. The update is conveyed to the clients, and the process is repeated until the model achieves the desired performance.

Compared to central approaches, FL addresses data privacy by enabling clients to keep their private data local and only send local model updates to the federator instead. However, the sensitive information contained in local model updates may facilitate model inversion attacks that allow the reconstruction of the clients' private data ~\cite{zhu2019deep,geiping2020inverting}. Private aggregation protocols ~\cite{bonawitz2017practical,aono2017privacy} ensure the privacy of individual updates by restricting the federator to only observe the aggregate of clients' updates.

Besides, FL suffers from security threats caused by Byzantine clients who deliberately manipulate the training through corrupt updates. 
The authors of \cite{blanchard2017machine} show the vulnerability of any simple linear aggregation rule such as FedAvg \cite{mcmahan2017communication} against Byzantine attacks. Even a single Byzantine worker can take full control of the learning process and make the model diverge or converge to a corrupt state. 
Countermeasures against such attacks, often in the form of robust aggregation rules, identify outliers in the clients' updates and exclude them from the aggregation. However, such techniques \cite{blanchard2017machine,yin2018byzantine,cao2020fltrust,zhao2022fedinv} require the federator to know the individual updates to compute statistics, which compromises the clients' privacy.

Thus, security requires access to individual local updates to learn statistics; whereas privacy requires concealing individual updates. This tension poses challenges for methods that jointly address security and privacy, which is the focus of this work.

Many Byzantine-resilient secure aggregation schemes \cite{so2020byzantine, jahani2023byzantine,velicheti2021secure,hao2021efficient,ma2022shieldfl,xhemrishi2025fedgt, gehlhar2023SAFEFL,dong2023privacy,hu2023efficient, zhong2024wvfl, zhang2024nspfl, lu2023robust,dong2021oblivious,miao2024rfed,egger2025byzantine} either are based on clustering and compromise privacy, require two federators, or consider computational privacy. Among computationally private schemes, the most closely related works are BREA~\cite{so2020byzantine} and ByzSecAgg~\cite{jahani2023byzantine}. By using secret sharing, the distance-based robust aggregation rule Krum~\cite{blanchard2017machine} is made privacy-preserving. Both schemes are computationally private and leak the pairwise distances between local updates to the federator. Additionally, BREA leaks extra information to the federator, through not re-randomizing secret sharings, cf. ~\cite{xia2024byzantine}. ByzSecAgg alleviates this problem by incorporating additional randomness.

While computational privacy has been the main focus in most studies due to its efficiency, all methods rely on computational hardness assumptions, potentially vulnerable to attacks such as ones run by quantum computers. If only a limited number of entities collude to compromise privacy, information-theoretic (IT) privacy provides the strongest possible privacy guarantee, even against computationally unbounded adversaries. In ByITFL~\cite{xia2024byzantine}, the authors proposed an information-theoretically private solution without privacy compromises that is based on the robustness of FLTrust ~\cite{cao2020fltrust}. However, ByITFL suffers from high communication and computation costs, growing with the fifth power of the number of clients. 
Currently, no Byzantine-resilient scheme in the literature provides both full IT-privacy and a practical communication cost.

We introduce \ourschemelo, a \underline{lo}w communication \underline{By}zantine-resilient and information-theoretically (\underline{IT}) private secure \underline{FL} aggregation scheme with total communication and computation cost quadratic in the number of clients. Similar to ByITFL, we utilize ideas from FLTrust \cite{cao2020fltrust} for Byzantine-resilience. The federator possesses a small representative root dataset to obtain a federator model update. This enables the computation of a trust score (TS) for each client through a discriminator function on the cosine similarity between the client's and federator's model update. %
A polynomial discriminator function enables IT privacy in \ourschemelo. We require a Trusted Third Party (TTP) during initialization to distribute Beaver triples ~\cite{beaver1992efficient} \emph{once} before the start of the training. Through an additively homomorphic secret sharing, each client's model update is shared with all other clients. 
Beaver triples enable multiplicative homomorphism, and hence, 
polynomial computations can be performed on shared model updates without increasing the degree of the encoding polynomial. 
Before reconstruction, an additively homomorphic Message Authentication Code (MAC) serves as an integrity check to prevent malicious clients from sending corrupt messages during the polynomial computation ~\cite{bendlin2011semi}. The federator decodes the aggregation and updates the global model. The proposed scheme is resilient against any $\Byzantine$ Byzantine clients, IT private against the federator and against any $\Colluding$ colluding clients, and robust against any $\Dropout$ dropouts, as long as $\Totalclient \geq \Byzantine+\Colluding+\Dropout+1$.

The concurrent work~\cite{ghavamipour2024privacypreserving} builds on similar methods and proposes a Byzantine-robust and privacy-preserving aggregation rule for decentralized learning without a federator. While~\cite{ghavamipour2024privacypreserving} targets computational privacy under a given number of collusions, we focus on the stronger IT privacy.

\section{System Model and Preliminaries}

\textbf{Notation.} We use $[n]$ to denote the set of positive integers $\{1,\cdots,n\}$ and use $\lfloor x \rfloor$ for the largest integer less than or equal to $x$. Vectors are denoted in bold type and scalars in normal type. $\langle \mathbf{x},\mathbf{y}\rangle$ denotes the dot product of $\mathbf{x}$ and $\mathbf{y}$. For an ordered sequence $x_1, \dots, x_n$ and a set $\mathcal{S} \subseteq [n]$ of indices, let $x_{\mathcal{S}} = \{x_i: i \in S\}$. The mutual information between two discrete random variables $X$ and $Y$ is denoted as $I(X;Y)$ while $H(X)$ denotes the entropy of $X$. 

\subsection{System Model} We consider FL with a federator, $\Totalclient$ clients, and a Trusted Third Party (TTP), which is used only before the start of the training process. The federator is honest-but-curious, $\Byzantine$ clients are Byzantine, $\Colluding$ clients are colluding, and $\Dropout$ clients may drop out during the execution. Each client $i$, $i \in [n]$, holds a private dataset $\Dataset_i$. The federator holds a small representative root dataset $\Dataset_0$ and coordinates the training process. It possesses a $\dimension$-dimensional global model $\model \in \mathbb{R}^\dimension$ and aims to find the best global model $\model^* = \operatorname{argmin}_{\model}F(\Dataset, \model)$ using private data held by the clients, where $F(\Dataset, \model)$ is the loss function over the global dataset $\Dataset=\cup_{i=1}^\Totalclient \Dataset_i$ .
In each global iteration $\globalIteration$, the federator sends the current global model $\model^{(\globalIteration)}$ to the clients. Client $i$ initializes its local model to the current global model, i.e., $\model^{(\globalIteration,0)}_i=\model^{(\globalIteration)}$, and updates it for $C$ local iterations by training on $\Dataset_i$:
\begin{equation*}
    \model^{(\globalIteration,\localIteration+1)}_i = \model^{(\globalIteration,\localIteration)}_i-\learningRate_i \cdot \nabla F(\Dataset_i; \model_i^{(\globalIteration,\localIteration)}), 
\end{equation*}
where $\learningRate_i$ is the local learning rate and $0<\localIteration \leq C-1$ is the local iteration. Each client $i$ sends the local model update $\modelupdate^{(\globalIteration)}_i=\model^{(\globalIteration,C)}_i-\model^{(\globalIteration)}$ to the federator. Meanwhile, the federator trains on $\Dataset_0$ and obtains $\modelupdate^{(\globalIteration)}_0$, which is assumed to be public. The federator aggregates the received local updates according to some aggregation rule $\AGG$, i.e., $\modelupdate^{(\globalIteration)} = \AGG(\modelupdate^{(\globalIteration)}_0,\modelupdate^{(\globalIteration)}_1, \cdots, \modelupdate^{(\globalIteration)}_n),$ 
and, using the global learning rate $\learningRate$, computes the global model for the next iteration\footnote{
Although we present our scheme in the simple gradient descent setting, it does not depend on the exact update rule and is applicable to, e.g., momentum and higher order methods and adaptive learning rate schedules. Similarly, it is compatible with additional privacy mechanisms like differential privacy~\cite{dwork2006differential}.}
\[\model^{(\globalIteration+1)} = \model^{(\globalIteration)}-\learningRate \cdot \modelupdate^{(\globalIteration)}. \]

\subsection{Preliminaries} 
We design a scheme that provides an \emph{information-theoretic} privacy guarantee, meaning the adversary's observations are shown to be (conditionally) statistically independent of the private data or, equivalently, the (conditional) mutual information between the observations and the private data equals zero. We condition on random variables the adversary \emph{must} have access to for the execution of the learning process.
We rely on \emph{linear secret sharing}. An $(\Totalclient,\Colluding)$-linear secret sharing (SS) scheme encodes a secret $s$ over a finite field $\mathbb{F}_p$ into $n$ secret shares $s[1], \cdots, s[n]$. Any $\Colluding$ or less shares are statistically independent of the secret (thus providing IT privacy with respect to $\Colluding$ colluders), while any $\Colluding+1$ or more shares may allow full reconstruction of the secret. The encoding is linear, i.e., homomorphic under addition and scaling.
A \emph{Beaver triple}~\cite{beaver1992efficient} consists of two independent and uniformly distributed random values $a$ and $b$ over a finite field $\mathbb{F}_p$ and their product $c=ab$, independently secret shared with $\Totalclient$ clients using a $(\Totalclient, \Colluding)$-linear secret sharing. Using a Beaver triple allows the clients to privately compute the multiplication of two secret shared values $x$ and $y$ using only additions and scaling by constants. Each client obtains a share of $xy$ and no information about $x$ or $y$ is leaked.

\subsection{Threat Model and Defense Goals}
Our attack model allows Byzantine clients to arbitrarily deviate from the protocol and access each client's dataset, for the purpose of choosing corrupt model updates. \ourschemelo is resilient and converges even if up to $\Byzantine$ clients collaboratively misbehave. 
\ourschemelo guarantees perfect IT privacy of honest clients' local model updates in each iteration against eavesdroppers with unlimited computing resources. Considering up to $\Colluding$ colluding clients and an honest-but-curious federator, we omit the notation of global iteration for brevity and adopt the privacy constraint from ~\cite{xia2024byzantine}.

\textit{Privacy against colluding clients:} 
Let $\mathcal{T}$ be the set of at most $\Colluding$ colluding clients, $\left | \mathcal{T} \right | \leq \Colluding$ and $\mathcal{E}$ be the set of at most $\Byzantine$ Byzantine clients, $\left | \mathcal{E} \right | \leq \Byzantine$. Let $M_{\mathcal{T}}$ denote the messages received by clients $i \in \mathcal{T}$, and $M_f$ denote the intermediate messages received by the federator. The colluding clients cannot learn any additional information about the honest clients beyond the current global model required for the learning tasks, and beyond what their local datasets and model updates leak about the honest clients' data, i.e.,
\begin{equation}\label{eq:ttp_privacy_colluding}
    I(\modelupdate_{[n] \setminus \mathcal{T} \cup \mathcal{E}}; M_{\mathcal{T}} \mid \modelupdate_0, \modelupdate_{\mathcal{T}}, \Dataset_{\mathcal{T}}, \model, \modelupdate) = 0.
\end{equation}

\textit{Privacy against the federator:} Knowing the current global model, the federator's local update and the root dataset, the federator should not gain any information about the local model updates of the honest clients beyond the aggregation, i.e.,
\begin{equation}\label{eq:ttp_privacy_federator}
    I(\modelupdate_{[n] \setminus \mathcal{T}\cup \mathcal{E}}; M_f \mid \modelupdate_0, \Dataset_0, \model, \modelupdate) = 0
\end{equation}

We consider delays or dropouts of up to $\Dropout$ clients during the protocol execution. Therefore, the protocol should be IT private against the honest-but-curious federator and against any collusion of up to $\Colluding$ clients, robust against $\Byzantine$ Byzantine clients, and tolerant to the to up to $\Dropout$ clients not responding during the execution.

\section{\ourschemelo} \label{sec:scheme}
We present \ourschemelo, a Byzantine-resilient scheme with IT privacy and low communication cost. The Byzantine robustness of \ourschemelo is based on the robustness of FLTrust~\cite{cao2020fltrust} and the integrity check of Message Authentication Codes. IT privacy comes through embedding each client's update into a finite field through stochastic quantization, and secret-sharing with all clients by linear SS schemes such as Shamir SS ~\cite{shamir1979share}, McEliece-Sarwate SS ~\cite{mceliece1981sharing}, Lagrange coded computing ~\cite{yu2019lagrange} and additive SS (see e.g. ~\cite{bendlin2011semi}). These linear SS schemes are inherently additive homomorphic, and multiplicative homomorphism can be unlocked through Beaver triples. 

We split the scheme into two phases: the initialization phase and the training phase. In the former, we assume the existence of a TTP for distributing Beaver triples to the clients. In the latter, per iteration, clients share their local model updates with other clients by linear SS. With Beaver triples converting multiplication of the secret shares to addition and scaling, the discriminator function, which is a polynomial function, can be computed on the secret shares to obtain the shares of the aggregation result. Note that, during the computation, to prevent Byzantine clients from providing corrupt computation results, we utilize additive homomorphic MACs for an integrity check. To this end, the TTP assigns a MAC to each secret share of the local model updates in the initialization phase. When performing computations on the secret shares, the clients have to perform the same computations on the corresponding MACs, and send the computation results and their MACs to the federator to prove the correctness of its computation. Therefore, the federator is able to identify misbehaving Byzantine clients and reconstruct the aggregation result correctly. %
We detail in the following the initialization phase and the training phase of \ourschemelo.

\subsection{Initialization Phase}
A TTP generates a sufficient number of Beaver triples $a,b,c$ and sends the corresponding secret shares to the clients, enabling multiplications on shares in the training phase. In addition, the TTP samples sufficiently many vectors $\randomvector^{(\globalIteration)}_{i} \in \mathbb{F}_p^\dimension$ and values $\random^{(\globalIteration)} \in \mathbb{F}_p$ uniformly at random. The TTP sends the random vectors $\randomvector^{(\globalIteration)}_{i}$ and the secret shares $\random^{(\globalIteration)}[i]$ to client $i \in [\Totalclient]$, one for each iteration $\globalIteration$, and the secret shares $\randomvector^{(\globalIteration)}_{i}[j]$ to client $j \in [\Totalclient]$. The random numbers are for secret sharing the local updates in the training phase efficiently and preserving the privacy of intermediate computations. We now focus on one particular iteration and omit $\globalIteration$ for brevity. 

To prevent Byzantine clients from providing corrupt computations, the TTP assigns an IT one-time MAC~\cite{bendlin2011semi} for each secret share distributed by the TTP: $\operatorname{MAC}_{\alpha, \beta}(x[i])=\alpha \cdot x[i]+\beta$, where $x[i]$ is the $i$th secret share of secret $x$, $(\alpha, \beta)$ is the MAC key and both $\alpha$ and $\beta$ are sampled uniformly at random from $\mathbb{F}_p$. We keep $\alpha$ globally fixed and sample a new $\beta$ independently for each MAC, thus, the MAC is additively homomorphic. This allows the integrity check for any linear computation $f$ on a secret share $x[i]$. Any linear computation $f$ performed on the $x[i]$ also has to be performed on the corresponding MAC, i.e., the clients compute $f(x[i])$ and $f(\operatorname{MAC}_{\alpha, \beta}(x[i]))$. The party holding the MAC keys checks the consistency of %
$f(\operatorname{MAC}_{\alpha, \beta}(x[i]))$ and the MAC of the computation $\operatorname{MAC}_{\alpha, \beta}(f(x[i]))$ (in the sequel we omit $\alpha$ and $\beta$ for brevity).
We use $\{ x[i] \}$ to denote the secret share $x[i]$ accompanied by the corresponding MAC, i.e. $\{ x[i] \}=\{ x[i], \operatorname{MAC}(x[i])\}$.
The TTP sends the MAC keys $(\alpha, \beta)$ used to generate the MACs for Beaver triples and random numbers to the federator in the initialization phase. The federator can then perform the integrity check at the end of each computation, mark clients who do not pass the check as Byzantine, and exclude them from future computations. In other words, at the end of the initialization phase, each client $i$ receives $\randomvector_{i}$, $\{\random[i]\}$, $\{ \randomvector_{j}[i] \},  j\in[\Totalclient]$ and a sufficient number of shares of Beaver triples $\{a[i]\}$, $\{b[i]\}$, $\{c[i]\}$, while the federator receives all MAC keys $(\alpha, \beta)$ used for generating MACs of these variables. The numbers of random numbers and Beaver triples required will be given later.

\subsection{Training Phase}
The training phase is conducted in five steps.

\textbf{Step $1$}: Clients and the federator \textbf{normalize and quantize} their model updates. Byzantine attacks can corrupt both the magnitude and direction of the model update vector. For defending magnitude attacks, updates $\modelupdate_i$, $\forall i \in \{0,\cdots,\Totalclient\}$, are normalized, i.e., the federator and the clients compute $\realnormalmodelupdate_i= \frac{\modelupdate_i}{\| \modelupdate_i \|}$. Since IT private SS scheme is performed over finite fields, we use an unbiased element-wise stochastic quantizer $Q_q(x)$ and a function $\phi(x)$, mapping integers to finite field elements, as in~\cite{so2020byzantine, jahani2023byzantine, xia2024byzantine}, with quantization step size $\frac{1}{q}$ and $2q+1$ quantization intervals. These functions map $\realnormalmodelupdate_i$ over the reals to a vector in prime field $\normalmodelupdate_i \in \mathbb{F}_p^\dimension$, where $p$ is a sufficiently large prime. The relation between $p$ and $q$ is detailed later.

\textbf{Step $2$}: The federator broadcasts $\normalmodelupdate_0$ to all clients. 
Each client $i$ \textbf{secret-shares} its \textbf{local update} $\normalmodelupdate_i$ with all clients using a linear SS scheme, thus keeping it private while making it computationally available.
The secret sharing of client $i$'s model update is supported by the uniformly random vector $\randomvector_i$ distributed in the initialization phase. Specifically, each client $j \in [\Totalclient]$ has a share $\randomvector_i[j]$ and client $i$ knows the plain value of $\randomvector_i$. Client $i$ secret-shares $\normalmodelupdate_i$ by broadcasting $\normalmodelupdate_i-\randomvector_i$ to all other clients.
Client $j$, for $j \in [n]$, can compute 
\begin{equation}
    \{ \normalmodelupdate_i[j] \} = (\normalmodelupdate_i-\randomvector_i) + \{ \randomvector_i[j]\}.
\end{equation}
Note that this constitutes an $(\Totalclient, \Colluding)$-linear SS of $\normalmodelupdate_i$ among clients $j\in [\Totalclient]$ as detailed in the following. 
The uniformly random vector $\randomvector_i$ is independent of $\normalmodelupdate_i$ and serves as a one-time-pad~\cite{shannon1949communication}. 
Once $\normalmodelupdate_i-\randomvector_i$ is public, gaining information about $\normalmodelupdate_i$ implies gaining information about $\randomvector_i$, which is only possible if the adversary has access to more than $\Colluding$ shares $\normalmodelupdate_i[j]$. Conversely, from any $\Colluding+1$ shares $\randomvector_i[j]$ it is straightforward to decode $\randomvector_i$ and thus $\normalmodelupdate_i$.
The corresponding $\operatorname{MAC}(\normalmodelupdate_i[j])$ is obtained by the additive homomorphism of the MAC. To summarize, each client $i$ computes the secret share and the corresponding MAC of the local model update of every client $j$, i.e. $\{ \normalmodelupdate_j[i] \}$, for $j \in [n]$. Thus, in each iteration, we need a total number of $\Totalclient$ 
vectors $\randomvector_{i}$ for this secret sharing.

\textbf{Step $3$}: To prevent Byzantine clients from incorrectly performing the normalization, each client $i \in [n]$, \textbf{validates the normalization} by computing secret shares of the squared $l2$-norm as
    $\{ (\| \normalmodelupdate_j \|_2^2) [i]\}= \{ \langle \normalmodelupdate_j[i], \normalmodelupdate_j[i] \rangle \},$ 
for the local model update $\normalmodelupdate_j$ of each client $j \in [n]$. Since $(\| \normalmodelupdate_j \|_2^2) [i]$ is a $\dimension$-dimensional vector, every $\{ (\| \normalmodelupdate_j \|_2^2) [i]\}$ computation is done by consuming $\dimension$ Beaver triples.
Specifically, a total of $\Totalclient$  $\dimension$-dimensional Beaver triples $\bm{o},\bm{v},w$ are required, where $\bm{o}, \bm{v}\in \mathbb{F}_p^\dimension$ and $w$ is the dot product of $\bm{o}$ and $\bm{v}$.
Clients send the computation results to the federator, who performs an integrity check on each $\{ (\| \normalmodelupdate_j \|_2^2) [i]\}$ using the MACs and reconstructs the value of $\| \normalmodelupdate_j\|_2^2 $. Similar to ~\cite{xia2024byzantine}, the federator checks if $\| \normalmodelupdate_j \|_2^2 $ lies within a certain interval, i.e., 
$\left |   \| \normalmodelupdate_j \|_2^2 - \phi(q\cdot Q_q(1))^2 \right | < \varepsilon \cdot q^2,$ 
where $\varepsilon$ is a predefined small threshold, to compensate for quantization errors. Clients who fail the normalization check are excluded from future computations.

\textbf{Step $4$}:  Clients apply the discriminator function on the secret shares of the model updates, and hence, obtain \textbf{secret shares of the aggregation result}. This step is based on the non-private scheme FLTrust~\cite{cao2020fltrust} to defend Byzantine attacks in direction. FLTrust assigns to each client a trust score determined by the cosine similarity between the federator and the local model update, i.e., $\operatorname{TS}_i = \operatorname{ReLU}(\cos(\theta_i))$, where $\theta_i$ is the angle between the updates and $\ReLU(x) = \max(0,x)$ is the rectified linear unit function. We call the function used to compute trust scores discriminator function, i.e., FLTrust uses ReLU as the discriminator. The federator then scales the local updates by their trust scores and aggregates them.

As discussed in~\cite{xia2024byzantine}, making FLTrust IT private is not straightforward and requires replacing ReLU as the discriminator function by a polynomial. The selection of the discriminator function will be discussed later. In general, the discriminator function is chosen to be a polynomial $h(x)$ with degree $\approdegree$, i.e. $h(x)=h_0+h_1x+ \cdots + h_kx^{\approdegree}$, providing a trust score for each client
$\trustscore_i = h(\cos(\theta_i))=h(\langle \normalmodelupdate_0,\normalmodelupdate_i \rangle), \forall i \in [\Totalclient].$
The aggregation result reads
\begin{equation}\label{eq:aggregate}
\begin{aligned}
    & \modelupdate =\frac{\| \modelupdate_0 \|}{\sum_{i\in [\Totalclient]}{\trustscore_i}} \cdot \sum_{i\in [\Totalclient]}{(\trustscore_i \cdot \normalmodelupdate_i)} = \| \modelupdate_0 \|\cdot \frac{\sumtwo}{\sumone}, \\
    & \text{where}\ \sumone = \!\sum_{i\in [\Totalclient]}{h(\langle \normalmodelupdate_0,\normalmodelupdate_i \rangle)} \ \text{and}\ \sumtwo = \!\sum_{i\in [\Totalclient]}{(h(\langle \normalmodelupdate_0,\normalmodelupdate_i \rangle) \cdot \normalmodelupdate_i)}.
\end{aligned}
\end{equation}
Since the federator holds $\modelupdate_0$, it suffices to compute the quotient $\sumtwo/\sumone$ in a privacy-preserving manner. 
Since the discriminator function is a polynomial, $\sumone$ and $\sumtwo$ are polynomial functions of the $\normalmodelupdate_i$, for $i \in [n]$. 
Having $\normalmodelupdate_0$, $\{ \normalmodelupdate_j[i] \}$ for $j \in [n]$, and $\random[i]$, client $i$ computes $\{(\random \sumone)[i]\}$ and $\{(\random \sumtwo)[i]\}$ by performing polynomial computations on $\normalmodelupdate_0$ and $\{ \normalmodelupdate_j[i] \}$ step by step. Note that, we require one $\random$ per iteration, whose shares are distributed by the TTP in the initialization phase. For each multiplication, the clients consume one Beaver triple. 
To be specific, in each iteration, for $h(\langle \normalmodelupdate_0,\normalmodelupdate_i \rangle)$ and $h(\langle \normalmodelupdate_0,\normalmodelupdate_i \rangle) \cdot \normalmodelupdate_i$ for $i \in [n]$, a total of $(\approdegree-1)\Totalclient$ scalar Beaver triples $a,b,c$ and $\Totalclient$ Beaver triples $x,\bm{y}, \bm{z}$ are needed, respectively, where $x\in \mathbb{F}_p$, $\bm{y} \in \mathbb{F}_p^\dimension$ and $\bm{z}\in \mathbb{F}_p^\dimension$ is the component-wise multiplication of $x$ and $\bm{y}$. The computations of $\random\sumone$ and $\random\sumtwo$ also require one scalar Beaver triple and one Beaver triple in the form of $x,\bm{y}, \bm{z}$.

Therefore, $\{(\random \sumone)[i]\}$ and $\{(\random \sumtwo)[i]\}$ can be computed by utilizing the additive homomorphism of both the linear SS scheme and the MAC. Finally, the clients obtain one secret share, each, of $\random \sumone$ and $\random \sumtwo$ and send them to the federator.

\textbf{Step $5$}:  The federator \textbf{reconstructs the aggregation result}. After receiving the $\{(\random \sumone)[i]\}$ and $\{(\random \sumtwo)[i]\}$ from client $i \in [n]$, the federator first checks the integrity of the computation results to ensure the correctness of each secret share, then reconstructs the value of $\random \sumone$ and $\random \sumtwo$ by Lagrange interpolation, as long as it receives at least $t+1$ correct computations. Note that unlike previous works, we do not rely on the decoding of error-correcting codes ~\cite{mceliece1981sharing}, which is expensive in computation and requires a higher total number of clients. Instead, the federator ensures the correctness of all computations through the integrity checks. The quotient $\sumtwo/\sumone$ is obtained without additional leakage since $\random$ is a uniformly random value unknown to all participants. The federator converts $\sumtwo/\sumone$ from the finite field back to the real domain, effectively performing the inverse of the quantizer $Q_q(x)$ and the mapper $\phi(x)$, and computes the global model for the next iteration.

To guarantee the correctness of the reconstructed results, none of the intermediate computations may cause a wrap-around in the underlying finite field. Since each coordinate of $\normalmodelupdate_i$ is in the range $-q$ to $q$ and each summation and multiplication during the polynomial computation can increase the range, we require $p \geq 2\Totalclient \dimension^k q^{2\approdegree+1}+1$, where $\Totalclient$ is the total number of clients, $\dimension$ is the dimension of the model updates, $q$ is the quantization parameter and $\approdegree$ is the degree of the discriminator function.

\begin{table*}[tb] %
\caption{Computation (Comp) and Communication (Comm) analysis for the total number of clients $\Totalclient$ and the dimension of the model updates $\dimension$. In ByzSecAgg and ByITFL $\partition$ determines how many sub-vectors the model updates are partitioned into. \vspace{-.35cm}}\label{tab:lo_complexity}
\centering
\begin{tabular}{c|cc|cc}
            & Per-client Comp & Per-client Comm & Federator Comp & Federator Comm \\ \hline
        BREA~\cite{so2020byzantine} & $O(dn\log^2n+dn^2)$ &  $O(dn+n^2)$ & $O((dn+n^3)\log^2n\log\log n)$ & $O(dn+n^3)$ \\ \hline 
        ByzSecAgg~\cite{jahani2023byzantine} & $O(\frac{d}{m}n\log^2n+\frac{d}{m}n^2)$ &  $O(\frac{d}{m}n+n^2)$ & $O((\frac{d}{m}n+n^3)\log^2n\log\log n)$ & $O(\frac{d}{m}n+n^3)$  \\ \hline
        ByITFL~\cite{xia2024byzantine} & 
$O((\frac{d}{\partition}n^3+n^4)\log^2n\log\log n)$ &  $O(\frac{d}{m}n^3+n^4)$ & $O((\frac{d}{m}n+n^2)\log^2n\log\log n)$ & $O(\frac{d}{m}n+n^2)$ \\ \hline
        LoByITFL & $O((d+l)n)$ &  $O((d+l)n)$ & $O((d+l)n^2\log^2n\log\log n)$ & $O((d+l)n^2)$ 
        \end{tabular}
\vspace{-0.3cm}
\end{table*}

\begin{figure}[!t]
    \centering
    \resizebox{.7\linewidth}{!}{
    \vspace{-0.5cm}
    \input{figures/comparison_discriminator}}
    \caption{Comparing different discriminator functions $h(x)$.}
    \vspace{-0.55cm}
    \label{fig:discriminator}
\end{figure}
\textit{Choice of the Discriminator Function.}
In FLTrust~\cite{cao2020fltrust}, the ReLU function is chosen as the discriminator function, whose usage in \ourschemelo poses challenges. The ReLU function requires a comparison, which is difficult to compute under IT privacy, i.e., it inherently leaks information about the relative direction of local updates with the federator update. %
The design of our discriminator function is based on the following core observation: It is possible to allow the function to take small non-zero values on the left tail, i.e., the interval between $-1$ and $0$. If the Byzantine workers present corrupt updates resulting in large negative cosine similarities, we invert the vector via a negative trust score. Hence, such a corrupt update cannot harm the learning process beyond what a corrupt update with a positive cosine similarity is capable of. 
Based on these observations, we find that it is not required to approximate the ReLU function by large-degree polynomials with high precision as in ~\cite{xia2024byzantine}, which is only satisfactory when the degree $\approdegree$ is greater than $6$; instead, we carefully choose a degree-$3$ polynomial that mimics ReLU in the negative half. In the positive half, we choose a more conservative shape than ReLU by giving higher trust scores to local model updates that strongly point in the same direction as the federator model update. Uncertain local model updates are attenuated more aggressively. This is very similar to \cite{lu2023robust}, where the authors use a quadratic function on the right, and a constant $0$ on the left. As a by-product, the degree of our discriminator polynomial is decreased significantly to $k=3$. The chosen degree-$3$ polynomial is given by $h(x) = 0.46897526 x^3 + 0.56578977 x^2 + 0.1860353 x + 0.01363545$, and plotted in Figure~\ref{fig:discriminator} with the degree-$6$ polynomial in~\cite{xia2024byzantine} and ReLU.

\section{Theoretical Analysis} \label{sec:theory}

The following Theorem~\ref{thm:priavcy_robustness} proves the robustness of \ourschemelo against Byzantine behavior during the protocol execution and the strong IT privacy guarantee achieved by our scheme. All proofs will be provided in an extended version of this paper. %

\begin{thm}
    \label{thm:priavcy_robustness}
    \ourschemelo with $\Totalclient \geq \Byzantine+\Colluding+\Dropout+1$ guarantees\begin{enumerate}[label=\arabic*), leftmargin=1.5em, itemsep=0pt, parsep=0pt, topsep=0pt]
        \item IT privacy against any $\Colluding$ colluding clients according to \eqref{eq:ttp_privacy_colluding} and against the federator according to \eqref{eq:ttp_privacy_federator}, and 
        \item Byzantine-resilience, i.e. it computes the correct aggregation result according to~\eqref{eq:aggregate} when up to $\Byzantine$ clients are Byzantine.
    \end{enumerate}
\end{thm}

The communication and computation costs are as follows.
\begin{proposition}
\ourschemelo requires a client to communicate $O((d+k)n)$ and the federator $O((d+k)n^2)$ scalars. The computation cost are $O((d+k)n)$ and $O((d+k)n^2\log^2n\log\log n)$, respectively.
\end{proposition}

\begin{remark} 
When choosing a linear SS scheme such as McEliece-Sarwate SS~\cite{mceliece1981sharing} or Lagrange coded computing ~\cite{yu2019lagrange}, that partition the model updates into $\partition$ smaller sub-vectors, like ByzSecAgg and ByITFL, the communication and computation complexity can be effectively optimized by reducing every $\dimension$ to $\dimension / \partition$.
\end{remark}

In Table \ref{tab:lo_complexity}, we compare the communication and computation complexity of \ourschemelo to previous solutions, with respect to $\Totalclient$, $\dimension$ and the partitioning parameter $\partition$ in ByzSecAgg ~\cite{jahani2023byzantine} and ByITFL ~\cite{xia2024byzantine}. At the cost of a trusted third party distributing random vectors and Beaver triples in the initialization phase, \ourschemelo has low communication and computation costs while preserving  IT privacy, improving practicality.

\section{Experiments}\label{sec:experiments}
We numerically demonstrate the convergence of \ourschemelo, under various attacks, and compare it to FedAvg and FLTrust. Our experiments are based on the implementation provided by ~\cite{gehlhar2023SAFEFL}.
We consider MNIST ~\cite{deng2012mnist} and CIFAR-10 ~\cite{krizhevsky2009learning} distributed across $\Totalclient=40$ clients. On MNIST we train a 
three-layer dense neural network, and on CIFAR-10 a convolutional neural network (CNN). ReLU and cross-entropy are used as activation and loss function, respectively. In each iteration, the clients randomly sample a minibatch of 64 samples from their local training dataset and perform local training on the minibatch. The learning rate is chosen to be $\learningRate=0.1$ for MNIST and $\learningRate=0.01$ for CIFAR-10. We set the number of local iterations to $C=1$. As in \cite{cao2020fltrust}, the size of the root dataset $\Dataset_0$ is $100$.
\begin{table*}[tb] %
\caption{We report the mean and standard deviation of the test accuracy (TA) with 10 runs on MNIST. For scaling attack, we report TA/attack success rate (ASR). FedAvg under no attack achieves $0.962\pm0.001$ for both the  i.i.d. and non-i.i.d. setting.}
\label{tab:experimentsMNIST}
\centering
\vspace{-0.2cm}
\begin{tabular}{c|c|c|c|c|c|c}
    
    & Aggregation & LF Attack & TM Attack & Krum Attack & \makecell{FLTrust/ \\ \ourschemelo Attack} &  \makecell{Scaling Attack \\ (TS/ASR)}\\ \hline
    \multirow{2}{*}{\makecell{i.i.d.}} & FLTrust & $0.933 \pm 0.008$ & $0.891 \pm 0.041$ & $\bm{0.930 \pm 0.005}$ & $0.908\pm0.012$ & $0.930\pm0.011$ / $\bm{0.617\pm0.366}$ \\
& \ourschemelo & $\bm{0.940 \pm 0.008}$ & $\bm{0.924 \pm 0.046}$ & $0.925 \pm 0.013$ & $\bm{0.932\pm0.012}$ & $0.942\pm0.010$ / $0.874\pm0.276$\\ \hline
    \multirow{2}{*}{\makecell{non-i.i.d.}} & FLTrust & $0.928 \pm 0.005$ & $\bm{0.916 \pm 0.013}$ & $\bm{0.934 \pm 0.004}$ & $0.919 \pm 0.020$ & $0.933\pm0.009$ / $\bm{0.671\pm0.321}$ \\
& \ourschemelo & $\bm{0.939 \pm 0.006}$ & $0.911 \pm 0.019$ & $0.933 \pm 0.008$ & $\bm{0.941 \pm 0.006}$ & $0.945\pm0.007$ / $0.947\pm0.061$\\
\end{tabular}
\end{table*}

We assume $25\%$ of the clients are Byzantine ($\Byzantine=10$) and perform a label-flipping (LF) attack, a scaling attack, and Fang's attack~\cite{fang2020local} optimized for the aggregation rules Krum~\cite{blanchard2017machine}, Trimmed Mean (TM)~\cite{yin2018byzantine}, FLTrust and \ourschemelo. We term them Krum attack, TM attack, FLTrust attack, and LoByITFL attack, respectively. The LF attack follows the same setting as in \cite{fang2020local}. The scaling attack is equivalently known as the backdoor attack~\cite{bagdasaryan2020backdoor}. We design Fang's attack on \ourschemelo by following the design of the adaptive attack in FLTrust ~\cite{cao2020fltrust} but set the discriminator function as the chosen degree-$3$ polynomial.
To model the data distribution, as in \cite{cao2020fltrust}, the clients are randomly partitioned into $10$ groups. A training sample with label $j$ is assigned to group $j$ with probability $\gamma>0$ and to any other group with probability $\frac{1-\gamma}{9}$. Data are uniformly distributed to each client within the same group. We set $\gamma=0.1$ for the i.i.d. setting (i.e., homogeneous data distribution) and $\gamma=0.5$ for the non-i.i.d. setting (i.e., heterogeneous data distribution). 
We let $q=1024$ and $\varepsilon=0.02$ for the normalization validation. 

We depict the average results and standard deviations across $10$ runs per setting. In Table~\ref{tab:experimentsMNIST}, it can be found that the resilience achieved by \ourschemelo is comparable to FLTrust, as neither outperforms the other across all attack scenarios. Hence, replacing the ReLU function with an appropriately tuned degree-$3$ polynomial does not introduce significant additional vulnerabilities. However, neither scheme is able to effectively defend against backdoor attacks. 
Similar results can be found for CIFAR-10 in Fig~\ref{fig:experimentsCIFAR}.

\begin{figure}[t]
  \centering
  \vspace{0.2cm}
  \hspace{-0.45cm}
    \centering
    \resizebox{.8\linewidth}{!}{
    \input{figures/cifar_iid_original}}
  \caption{Testing accuracy of \ourschemelo on i.i.d. CIFAR-10 over iterations. For scaling attack, the final attack success rate is $0.922\pm0.012$ for LoByITFL and $0.920\pm0.009$ for FLTrust.}
  \label{fig:experimentsCIFAR}
  \vspace{-0.5cm}
\end{figure}

\section{Conclusion}
We introduced \ourschemelo, an IT private and Byzantine-resilient federated learning scheme that significantly reduces the communication cost of \cite{xia2024byzantine} towards practical applications by the use of a TTP during the initialization phase. We studied discriminator functions different from the originally proposed ReLU \cite{cao2020fltrust} that allow for IT private schemes without sacrificing the resiliency against Byzantine attacks. We prove the robustness and the privacy of our scheme, and experimentally compare the Byzantine-resilience of \ourschemelo against FLTrust \cite{cao2020fltrust}. While both schemes can effectively defend against LF attacks, TM attacks, Krum attacks, and Fang's attacks individually tailored to the choice of the discriminator polynomial, we found that none of the schemes can defend against a well-executed backdoor (scaling) attack. Improving our IT private scheme to also resist backdoor attacks while maintaining the strongest possible notion of privacy is left for future work.

\bibliographystyle{unsrt}
\bibliography{refs.bib}

\end{document}